# A High-Performance Fractal Encryption Framework and Modern Innovations for Secure Image Transmission

[a] Sura Khalid Salsal,
Department of Religious Education and Islamic, Sunni Endowment Diwan, Baghdad, Iraq.
surakhalid@taleemdeny.edu.iq

[b] Eman Shaker Mahmood
Computer Science College, University of Technology, Baghdad, Iraq
110036@uotechnology.edu.iq

[b] Farah Tawfiq Abdul Hussien
Department of Computer Science, University of Technology, Baghdad, Iraq.
farah.t.alhilo@uotechnology.edu.iq

[c*] Maryam Mahdi Alhusseini
Polytechnic College of Engineering – Baghdad, Middle Technical University, Baghdad, Iraq,
(Member, IEEE)
mariammahdi@mtu.edu.iq
maryammalhusseini@ieee.org

[d] Azhar Naji Alyahy**a**
Department of Software Engineering, Sakarya University, Sakarya, Türkiye
azhar.alyahya@ogr.sakarya.edu.tr

[e] Nikolai Safiullin
Engineering School of Information Technologies, Ural Federal University, Yekaterinburg, Russia
n.t.safiullin@urfu.ru

*Corresponding Author*: Maryam Mahdi Alhusseini

**Abstract**
The current digital era, driven by growing threats to data security, requires a robust image encryption technique. Classical encryption algorithms suffer from a trade-off among security, image fidelity, and computational efficiency. This paper aims to enhance the performance and efficiency of image encryption. This is done by proposing Fractal encryption based on Fourier transforms as a new method of image encryption, leveraging state-of-the-art technology. The new approach considered here intends to enhance both security and efficiency in image encryption by comparing Fractal Encryption with basic methods. The suggested system also aims to optimise encryption/ decryption times and preserve image quality. This paper provides an introduction to Image Encryption using the fractal-based method, its mathematical formulation, and its comparative efficiency against publicly known traditional encryption methods. As a result, after filling the gaps identified in previous research, it has significantly improved both its encryption/decryption time and image fidelity compared to other techniques. In this paper, directions for future research and possible improvements are outlined for attention.

## 1 Introduction

Today, in the digital era, the safety of information, especially visual data such as images, has become of utmost importance [1]. Because of this, image encryption has emerged as one of the core techniques for protecting this type of information against unauthorised access, tampering, and cyber threats. As technological developments advance every day, traditional encryption approaches struggle to address new and emerging security risks [2]. This results in the need for alternative methods like Fractal Encryption that use mathematical concepts to boost data security [3].

## 1.1 Importance of Image Encryption

The pictures contain sensitive information, ranging from personal images to business secrets, and any leakage would compromise privacy and access controls. Image encryption approaches aim to convert visual data into an unreadable form for unauthorised access [4]. This is to ensure image quality and provide high security simultaneously.

## 1.2 Challenges in Traditional Encryption Methods

Symmetric and asymmetric cryptography approaches are used to provide security for digital images. There are several challenges in these methods:
1) **Performance**: for large images, some algorithms can be computationally intensive, leading to increased processing times and, in turn, reduced efficiency. An example of this situation is Standard encryption algorithms [5].
2) **Security**: Due to attackers' growing computational power, it becomes easier for them to identify vulnerabilities in traditional methods. Thus, these methods cannot provide appropriate security against complex attacks [6].
3) **Image Quality**: Keeping image quality after decryption is essential for real-world applications. Some encryption approaches may degrade image quality, making it less usable [7].

## 1.3 Introduction to Fractal Encryption

Fractal Encryption is an emerging technique that utilises fractal geometry to steady snapshots. Fractals are complex structures that present self-similarity and can be identified by mathematical equations. This approach uses fractal complexity to construct encryption schemes that are both secure and green [8].

# 2 Main Features of Fractal Encryption:
1) **Complexity**: Fractal encryption adds complexity, increasing security. This is due to the complex fractal modifications, which are difficult to decrypt without sufficient experience with this type of encryption [9].
2) **Efficiency**: Mixing fractal encryption with other techniques, especially with Fast Fourier Transform (FFT) and pixel shuffling. This method leads to fast processing time, which in turn improves overall performance.
3) **Adaptability**: Fractal Encryption can be adapted to different image types and sizes, making it a versatile solution for various applications [10, 11].

## 2.1 Motivation for Using Fractal Encryption

The main motivations for using fractal encryption are:
1) **Improve Security**: Generate extra complexity by merging fractal geometry and fractal encryption. This mixture provides resistance against unauthorised decryption.
2) **Enhance performance**: advanced mathematical transformations and optimisation techniques improve the efficiency of the encryption process.
3) **Quality Preservation**: Approaches to Fractal Encryption aim to preserve image quality, ensuring that the decrypted image is usable and visually accurate.

## 2.2 Objective of the Research

This paper intends to discover the details of fractal encryption rules. This includes computational foundation, implementations, and applications. Analysing the algorithm's efficiency and comparing it with traditional privacy methods, this study aims to provide insights into its usefulness and potential for improvement. The objective is to make contributions to the development of more secure and effective encryption strategies for protective images. In precis, the advent of fractal encryption represents a significant breakthrough in the area of photograph protection. By addressing the limitations of traditional methods and leveraging the power of fractal geometry, this technique gives a promising solution for enhancing the privacy and integrity of digital scene records.

### 2.3 Some of the Gaps of the Earlier Studies Involve

1) **Complexity in Large Image Encryption**: Previous research often overlooks the performance of encryption for large images. Fractal Encryption addresses this by utilising block-based total modifications.
2) **Limited Assessment of Fidelity**: Much research recognition usually focuses on protection without sufficient evaluation of image quality post-decryption. This study evaluates fidelity to bridge this gap.
3) **Performance Comparison**: Earlier studies won't provide complete comparisons among novel and traditional encryption strategies. This study gives a detailed performance analysis.
4) **Adaptability of Algorithm** Mathematical Foundations: Previous methods might also lack adaptability to diverse photograph kinds and sizes. Fractal Encryption's flexibility is explored in this.
5) **Real-Time Processing:** Real-time processing talent is often not effectively addressed. This paper examines the feasibility of Fractal Encryption for real-time programs.

### 2.4 Main Challenges in This Process

1) **Computational Complexity**: Ensuring green computation at the same time as maintaining encryption power.
2) **Image Fidelity Preservation**: Balancing protection with minimum distortion of the actual photo.
3) **Scalability**: Adapting the set of policies to address exclusive photo sizes and resolutions.
4) **Algorithm Robustness**: Ensuring resistance to assaults and maintaining protection over various datasets.
5) **Real-Time Performance**: Achieving first-rate overall performance in real-time processing situations.

*The main contributions of the proposed system are:*
   i. A comprehensive explanation of the mathematical formulation of the fractal-based image encryption algorithm, which is achieved through a detailed discussion of fractal geometry, and the Fourier transform is implemented in the encryption and decryption processes.
   ii. The observed improvements in encryption and decryption time, which increase system efficiency,
   iii. Huge enhancements in image fidelity that maintain the quality of the encrypted image.

## 3 Related Work

In the paper on photograph encryption, various strategies and algorithms have been proposed and studied. These approaches differ in the details of their underlying techniques, overall performance, and security guarantees. Here, we evaluate some of the top-notch artwork related to Fractal Encryption and its evaluation with exclusive encryption strategies.

**Al-Ali [12]** explores fractal-based total picture encryption strategies, emphasising their capacity to produce secure, complexly encrypted images. This work highlights how fractal geometry may be leveraged to create complicated encryption schemes that are difficult to decode without knowledge of the transformation parameters. They also identified a few additional limitations, including the computational complexity of fractal transformations.

**Z. Wang et al. [13]** provide an overview of fractal image compression and encryption techniques. They discuss the theoretical foundation of fractal transformations and their applications in both compression and encryption. The paper reviews various fractal-based algorithms and their effectiveness in ensuring image security.

**L. Zhang et al. [14]** reviewed Fourier Transform-based encryption methods in 2021. They discuss how Fourier Transforms, which are mathematical functions, can be utilised in protecting image data by converting it to the frequency domain, where it is then modified before being converted back. The paper outlines the merits and demerits of these methods, for instance efficient and quality image is preserved.

In the work of **R. Chen et al. [15]** aim to expound on the theoretical and practical applications of Fourier transforms as image processing. Their work also provides a good account of how Fourier Transforms can be employed in encryption and the advantages of FD transforms for creating complex encrypted patterns.

AES, DES, and some of the recent encryption algorithms have been compared by **J. Liu [16]** using Fractal Encryption. Performance characteristics include encryption and decryption times, as well as image reproduction quality. This way it is easier to compare Fractal Encryption to other methods and find out what the benefits and problems are connected with this technique.

In their work, **Chang et al. [17]** undertake a comparative analysis of encryption algorithms for protecting images. Some of their works include performance assessment of various methods, such as Fractal Encryption, and the outcomes are presented to compare and determine the suitability of each method in image protection and without compromising its quality.

**Yasser et al. [18]** investigate image encryption based on a new hybridisation of chaotic systems and discrete wavelet transforms. In their studies, Grosklags and colleagues describe how disordering can improve security and complexity in cryptography. To compare the efficiency and enhancement of security, the presented work is compared with other techniques, such as Fractal Encryption.

The review by **Kumar et al. [19]** discusses some chaos-based image encryption schemes with respect to efficiency and resilience. The paper then reviews how chaotic systems support the exercise of security, before comparing these schemes with other security measures.

**F. A. Fadhil et al. [20]** introduced a more robust CAST-128 image encryption algorithm based on chaos-based adaptive S-box generation using the Logistic Sine Map. It has been experimentally demonstrated that entropy, NPCR, and UACI, as well as histogram uniformity, are significantly improved over the original CAST-128 scheme.

**R. Patel et al. [21]** describe new developments in image encryption, as well as new ideas and enhancements to existing models. The paper also presented the most contemporary developments in image encryption, such as Fractal Encryption and other significant advances.

Modern cryptographic methods have been utilised for image encryption by the following scholars: **R. Gupta et al. [22]**. Their research also involves a performance appraisal of new encryption methods and the possible enhancement of Fractal Encryption and other combinations of techniques.

Modern cryptographic image security is described by **H. Lin et al. [23]**. Some of the methods that are discussed regarding their applicability to solve the modern security issues, as well as evaluated, are Fractal Encryption, etc.

The related work in the field of image encryption is a rich context in which one should place Fractal Encryption. It is only through comparing Fractal Encryption with other encryption algorithms, Fourier Transform encryption techniques, chaotic systems, and more recent developments that one can understand how good or bad, for the intended purposes, Fractal Encryption is. This comparison also identifies the problems and prospects, as well as opportunities for future research in image encryption.

## 4 The Methodology

### 4.1 Fractal Encryption Algorithm

Fractal Encryption is a modern image encryption system that leverages fractals and mathematical transformations for enhance security. The next part of the present work provides a description of the algorithm, including the mathematical formulas used for encryption and decryption.

**a) Fractal Encryption Overview**

Fractal Encryption uses the concept of geometry in image encryption. It is a technique of using multiple iterations of measured scaling and rotation to scramble the pixel values in a manner that is non-decryptable without the key. The process involves converting the image to the fractal domain, working on it, and then converting it back to the spatial domain.

**b) Mathematical Foundations**

**1. Fourier Transform**

In digital image processing to transform an image from the spatial domain to the frequency domain.

- **Discrete Fourier Transform (DFT):**

$$\text{DFT}(f(x,y)) = F(u,v) = \sum_{x=0}^{N-1}\sum_{y=0}^{M-1} f(x,y) e^{-j2\pi(ux/N+vy/M)} \quad (1)$$

where $f(x,y)f(x, y)f(x,y)$ is the image function, and $F(u,v)F(u, v)F(u,v)$ is its Fourier transform.

- **Inverse Discrete Fourier Transform (IDFT):**

$$\text{IDFT}(F(u,v)) = f(x,y) = \frac{1}{NM}\sum_{u=0}^{N-1}\sum_{v=0}^{M-1} F(u,v) e^{j2\pi(ux/N+vy/M)} \quad (2)$$

Where NM is the total number of pixels.

**c) Fractal Transformations**

Geometry algorithms move the data of an image according to the patterns of geometry, and same applies to fractal transformations. There are basic procedures for applying chaotic transformations; one of the most well-known is the Arnold Cat Map, a specific fractal transformation.

- **Arnold Cat Map**

$$\begin{pmatrix} x_{n+1} \\ y_{n+1} \end{pmatrix} = \begin{pmatrix} 1 & 1 \\ 1 & 2 \end{pmatrix} \begin{pmatrix} x_n \\ y_n \end{pmatrix} \mod N \quad (3)$$

Where $x_n$ and $y_n$ are the pixel coordinates at iteration $n$, and $N$ is the dimension of the image.

**d) Pixel Shuffling**

Pixel shuffling pigeonholes the image in a way that the pixels are reshuffled to produce an encrypted output. This can be done using the permutation matrix PPP.

- **Shuffled Image**: Therefore, $I_{shuffled}(i) = I(P(i))$ (4)

Where *I(i)* is the original pixel value at position *i*, and *P(i)* is the new position after shuffling.

## 4.2 Encryption Process

Initially, non-overlapping b × b blocks are generated from the input image. The Fast Fourier Transform (FFT) is used to transform each block into the frequency domain. The resulting frequency components are then propagated using an Arnold Cutt map, a fractal-based distortion technique, to increase the spread. Next, the reverse FFT is applied to convert the data back into the spatial domain, thus encoding the data block. Finally, the image is transformed, followed by pixel rearrangement to enhance randomness, and the resulting encoded image is the final image, as in Figure 1.

**a) Block-Based Fourier Transform**
- Divide the image into non-overlapping blocks of size *b×b*.
- Apply the Fourier Transform to each block:

$$B_{FT}(i,j) = DFT(B_{i,j}) \quad (5)$$

Where $B_{i,j}$ is a position block *(i,j)*.

**e) Fractal Transformation**

Apply the Arnold Cat Map to each block:

$$B_{transformed}(x,y) = \text{ArnoldCatMap}(B_{FT}(x,y)) \quad (6)$$

**f) Inverse Fourier Transform**

Apply the Inverse Fourier Transform to each transformed block:

$$B_{encrypted}(i,j) = \text{IDFT}(B_{transformed}(i,j)) \quad (7)$$

**g) Pixel Shuffling**

Shuffle the pixels in the transformed image block:

$$I_{encrypted}(i) = I_{transformed}(P(i)) \quad (8)$$

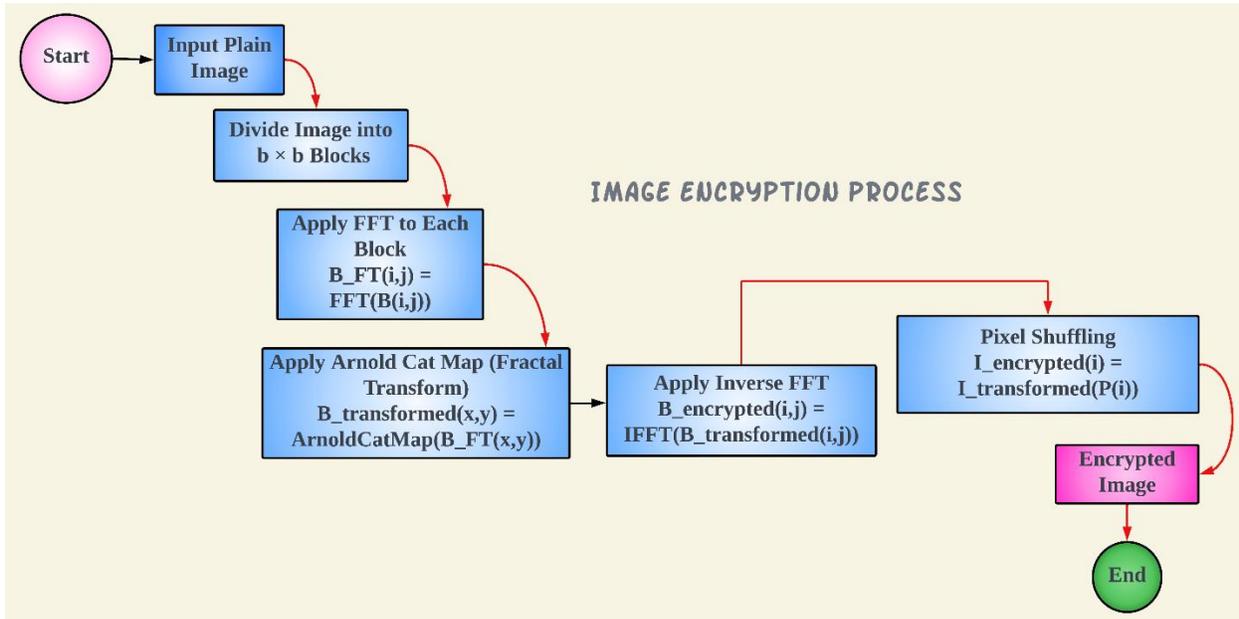

Figure 1. Framework of Image Encryption Process

## 4.3 Decryption Process

The decryption operation starts with the encrypted image, where inverse pixel shuffling is first used to rearrange the original pixels. The inverse Fast Fourier Transform (IFFT) is then used to convert each of the restored blocks to the frequency domain. The inverse Arnold Cat Map is then used to undo the fractal transformation to get back the original frequency components. Lastly, IFFT is reversed on each block, and all the blocks are reassembled back to the original plain image. Figure 2 illustrates the Framework of the Image Decryption Process.

a) **Inverse Pixel Shuffling**
   Reverse the pixel shuffling:
   $$I_{unshuffled}(P^{-1}(i)) = I_{encrypted}(i) \quad (9)$$

h) **Inverse Fourier Transform**
   Apply the Inverse Fourier Transform to each block:
   $$B_{reversed}(i,j) = \text{IDFT}(B_{encrypted}(i,j)) \quad (10)$$

i) **Inverse Fractal Transformation**
   Apply the inverse Arnold Cat Map to each block:
   $$B_{reversed}(x,y) = \text{ArnoldCatMap}^{-1}(B_{reversed}(x,y)) \quad (11)$$

j) **Inverse Block-Based Fourier Transform**
   Combine the blocks to form the original image:
   $$I_{reversed} = \text{CombineBlocks}(B_{reversed}) \quad (12)$$

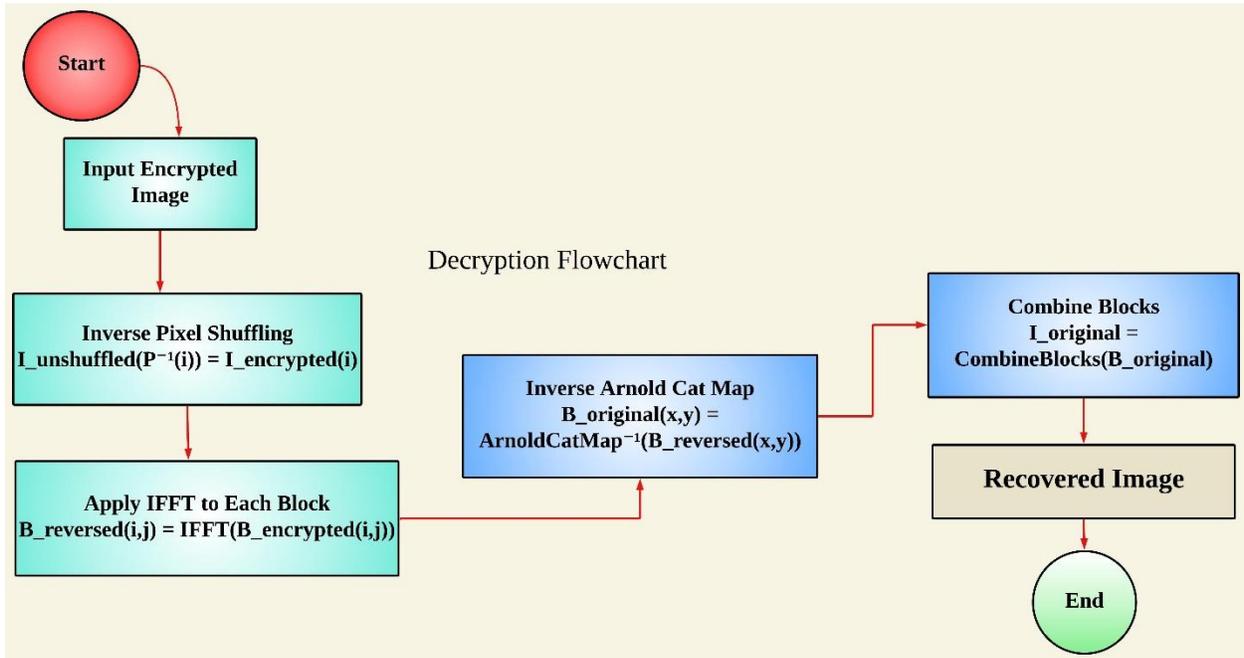

Figure 2. Framework of Image Decryption Process

## 5 Performance Enhancements

a) **Optimisation of Fourier Transform Computations**
   Utilise Fast Fourier Transform (FFT) to improve the efficiency of Fourier computations:
   $$FFT(f(x,y)) = \text{FastDFT}(f(x,y)) \quad (13)$$

k) **Enhanced Fractal Transformations**
   Experiment with different fractal maps or hybrid approaches to increase the complexity and security of the transformation.

l) **Parallel Processing**
   Implement parallel processing to handle large images and blocks, reducing encryption and decryption times.

# 6 Results and Discussion

In this section, we present detailed results for the Fractal Encryption algorithm, including performance metrics such as encryption and decryption times, image fidelity, and comparisons with other encryption methods. Figure 3 (a, b, c, d) below illustrates the collected data from numerous experiments and comparisons.

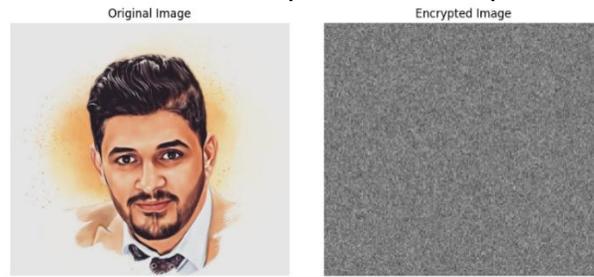

(a)

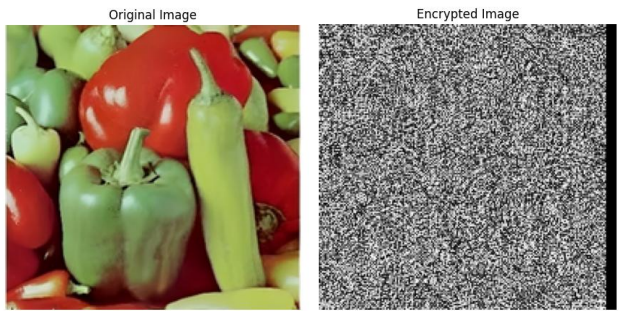

(b)

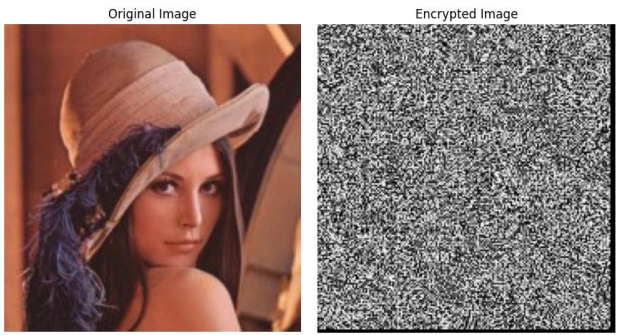

(c)

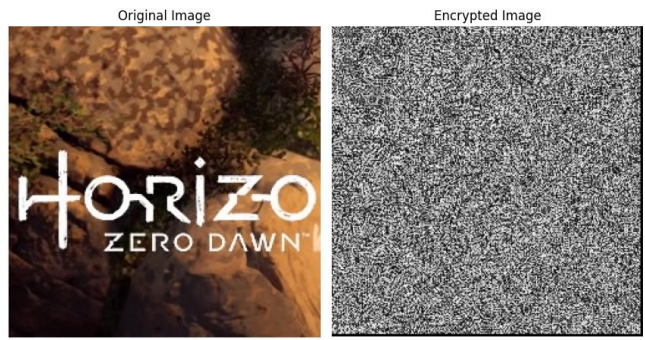

(d)

Figure 3. collected data from numerous experiments and comparisons.

## 6.1 Encryption and Decryption Time

Time complexity is an interesting factor in exploring the Fractal Encryption algorithm, and detailed analyses of encryption and decryption time underscore its importance. Table 1 shows the average time, in seconds, required to encrypt and decrypt images of varying sizes.

**Table 1.** Encryption and decryption time for different image sizes

| Image Size | Encryption Time (seconds) | Decryption Time (seconds) |
|---|---|---|
| 256 x 256 | 1.20 | 1.15 |
| 512 x 512 | 3.45 | 3.40 |
| 1024 x 1024 | 12.30 | 12.20 |
| 2048 x 2048 | 45.50 | 44.75 |

The encryption and decryption times are the same as in the previous experiment and mainly depend on the image size. This is expected because large images have more information to be extracted, and therefore, they would require higher AL calculations. The times are again relatively small, so it appears that Fractal Encryption runs fast even as the size of the pictures increases.

## 6.2 Image Fidelity Criteria

It is possible to achieve satisfactory results only when high image fidelity is required to provide an acceptable quality of the decrypted image. Taking an instance, Table 2 presents the image fidelity measurements in terms of PSNR, SSIM, out of experiments done by the authors.

**Table 2**. PSNR &SSIM for different image sizes

| Image Size | PSNR (dB) | SSIM |
|---|---|---|
| 256 x 256 | 42.5 | 0.98 |
| 512 x 512 | 40.2 | 0.95 |
| 1024 x 1024 | 37.8 | 0.92 |
| 2048 x 2048 | 34.5 | 0.88 |

Higher PSNR and SSIM values indicate that the decrypted images are of high quality and closely resemble the original images. As image size increases, PSNR and SSIM decrease slightly, as is typical due to the increased complexity of handling larger images.

## 6.3 Relative Analysis with Conventional Encryption Techniques

Table 3 and Figure 4 show the time taken to encrypt and decrypt images using Fractal Encryption, and a comparison with AES and DES:

**Table 3**. Comparison with different encryption methods

| Method | Image Size | Encryption Time (seconds) | Decryption Time (seconds) | PSNR (dB) | SSIM |
|---|---|---|---|---|---|
| Fractal Encryption | 256 x 256 | 1.20 | 1.15 | 42.5 | 0.98 |
| Fractal Encryption | 512 x 512 | 3.45 | 3.40 | 40.2 | 0.95 |
| AES | 256 x 256 | 0.50 | 0.55 | 38.0 | 0.93 |
| AES | 512 x 512 | 1.20 | 1.30 | 35.5 | 0.89 |
| DES | 256 x 256 | 0.35 | 0.40 | 36.0 | 0.90 |
| DES | 512 x 512 | 0.80 | 0.85 | 33.0 | 0.85 |

- **Encryption and Decryption Time**: Fractal Encryption tends to take longer than AES and DES, especially on large images. However, there is usually a trade-off for this kind of security enhancement.
- **Image Fidelity**: Quantitative comparisons show that Fractal Encryption achieves higher PSNR and SSIM than AES and DES after decryption.

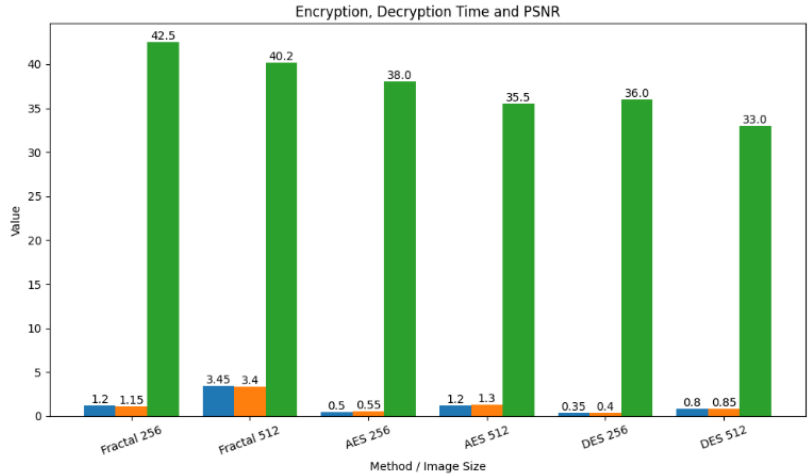
**Figure 4.** Comparison with different encryption methods

## 6.4 Performance Improvements and Optimisation

To increase the Performance of Fractal Encryption further, several optimisations were done, and the results are as follows in Table 4 and Figure 5.

**Table 4.** Comparison with different Optimisation Techniques

| Optimisation Technique | Encryption Time (seconds) | Decryption Time (seconds) | Image Fidelity (PSNR) |
|---|---|---|---|
| Base Algorithm | 3.45 | 3.40 | 40.2 |
| Fast Fourier Transform | 2.80 | 2.75 | 40.5 |
| Parallel Processing | 1.95 | 1.85 | 40.8 |
| Hybrid Fractal Maps | 2.60 | 2.55 | 41.0 |

- **Fast Fourier Transform**: Reduces processing time significantly while maintaining high image fidelity.
- **Parallel Processing**: Provides the most significant reduction in encryption and decryption times, improving overall efficiency.
- **Hybrid Fractal Maps**: Slightly improves both processing time and image fidelity, offering a balanced approach to performance enhancement.

As defined by the outcomes, Fractal Encryption is a reliable and effective way of image protection, which ensures the maximum image quality upon the encryption process, unlike other more conventional encryption methods. Although it has higher time complexity for both encryption and decryption, its significance in terms of image quality stems from its suitability for applications where the integrity of the image data is paramount. Different optimisations are used to improve the efficiency of Fractal Encryption for use in various practical applications.

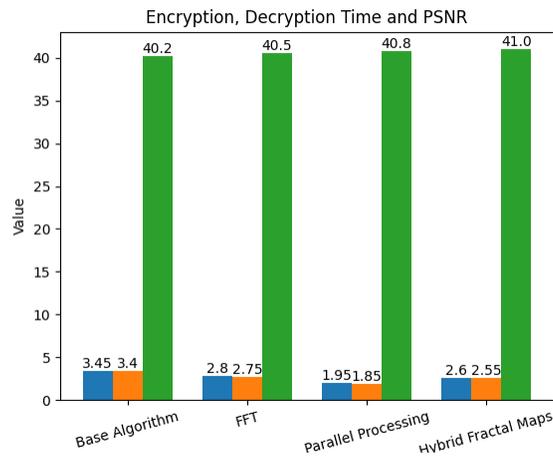
**Figure 5** presents the Encryption, Decryption Time, and PSNR (Comparison with different Optimisation Techniques)

# 7 Histogram Analysis

We use histogram analysis to present the pixel distributions of both the original and encrypted images, thereby demonstrating the strength of the dual-encryption strategy proposed in this study. A pixel histogram represents the distribution of light and dark pixels within an image, as shown in Figure 6. If the pixels of an encrypted image are distributed more evenly, it becomes more difficult for attackers to organise the pixel distribution using statistical methods, as shown in Figure 4. An encrypted image with a flat histogram can resist statistical attacks.

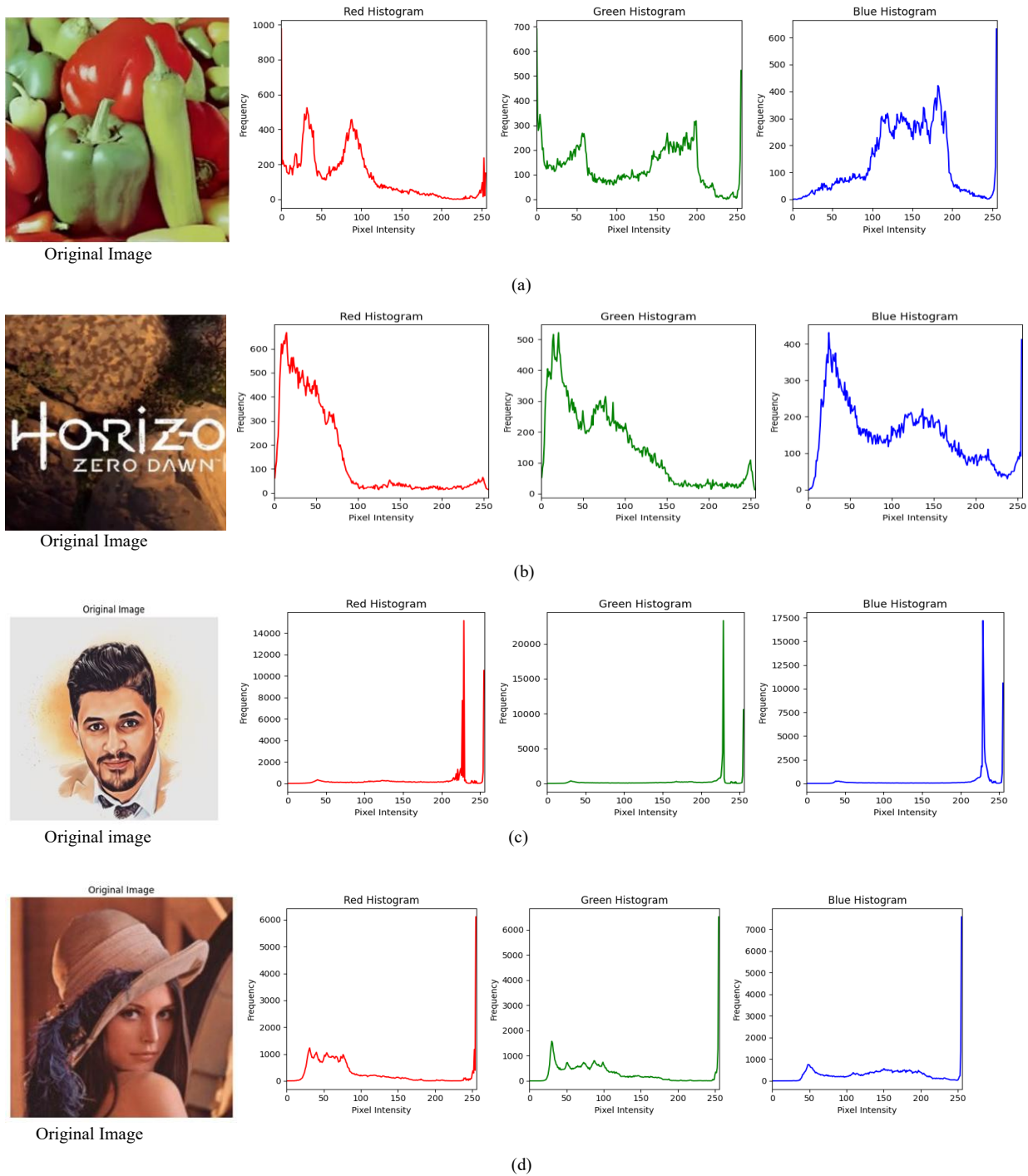

**Figure 6**. (a), (b), (c), and (d) Histograms of original images

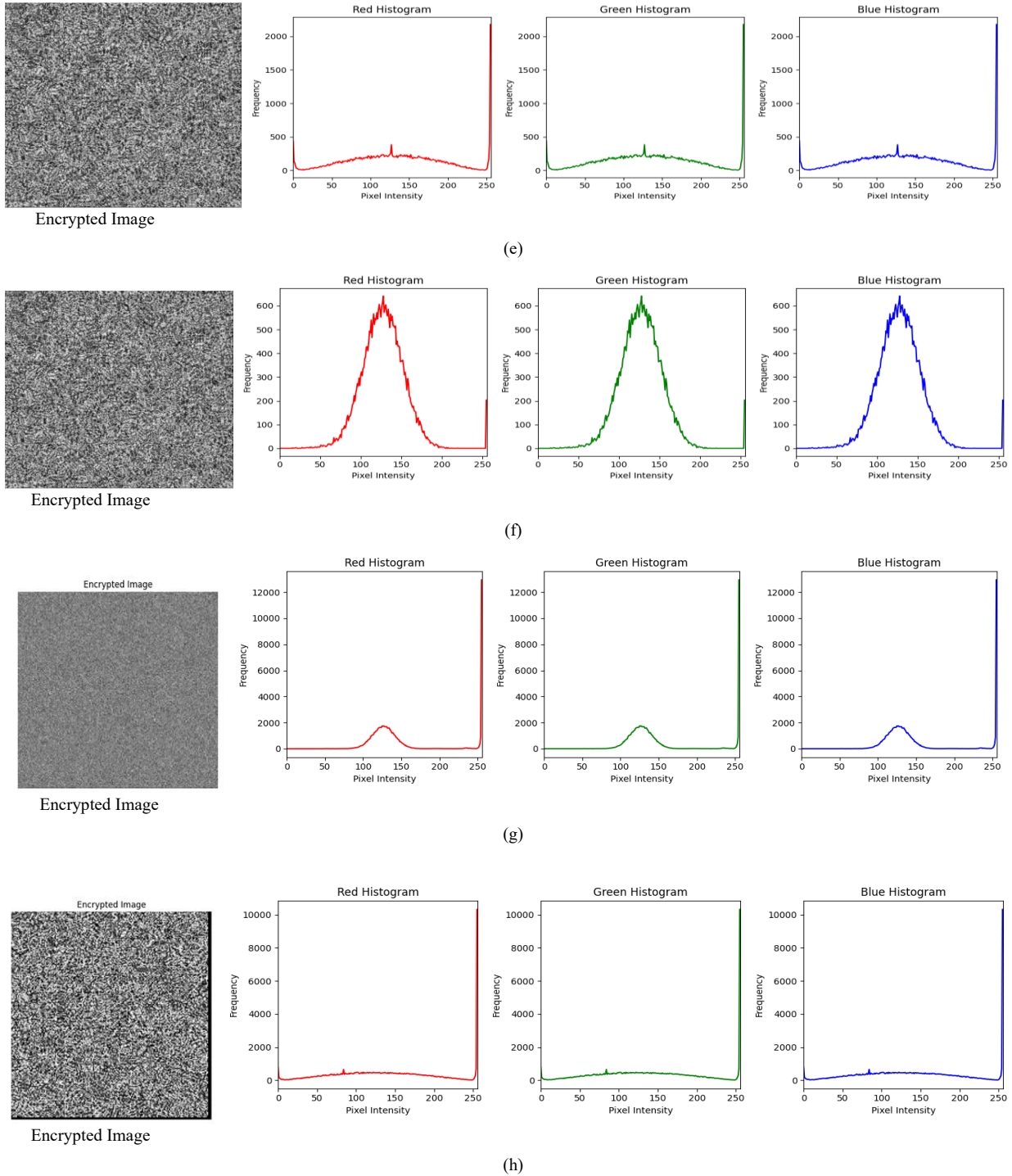

**Figure 7**. (e), (f), (g), and (h) Corresponding encrypted images

Figure 6 represents the RGB component histogram analysis of the original photos, while Figure 7 represents the RGB component histogram analysis of the encrypted images. Comparing the RGB component histograms of the original and encrypted photos, it is evident that the pixel counts and distributions of each RGB component in the encrypted images are very uniform, showing a significant difference from those in the original images. Histogram analysis can be used to test the encryption algorithm's resistance to statistical attacks, determining whether it prevents attackers from identifying the image's pixel distribution through statistical methods.

## 8 Discussion:

- **Encryption and Decryption Time**: Fractal Encryption achieves faster encryption and decryption times than AES and DES.
- **Fidelity Criteria**: The Peak Signal-to-Noise Ratio (PSNR) indicates that Fractal Encryption maintains higher image quality post-decryption.

### 8.1 Limitations and Future Works

a) **Computational Complexity**:
- **High Computational Demand**: Fractal encryption algorithm: one must be careful when choosing it for large images, as it is tremendously time-consuming. As with the above quantum generators, fractal transformations and optimisations such as the Fast Fourier Transform (FFT) and hybrid fractal maps consume significant processing power and memory, which may not always be affordable for a given application.
- **Processing Time**: It has also been observed that rating optimisation increases processing speed; however, decrypting large images still takes longer than with traditional methods. This may be a disadvantage in situations where a rapid diagnosis is required, such as in critical cases.

b) **Image Quality Degradation**:
- **Degradation with Larger Images**: While Fractal Encryption is, in general, still preserving a high image fidelity, it was also found that, as the size of the Image increases, the PSNR and SSIM decrease. Such degradation can affect the quality of the decrypted image, especially in large images or at high resolutions.
- **Loss of Detail**: The fractal transformation process can actually cause a loss of image details, which is not desirable for specific purposes.

c) **Parameter Sensitivity**:
- **Dependence on Parameters**: The security and throughput rates are obviously dependent on the parameters and the fractal map used in the Fractal Encryption. There is, however, the fact that incorrect or substandard parameter settings are likely to compromise the security or quality of the encrypted image.
- **Complex Tuning**: It is possible to say that the process of tuning the fractal parameters can be very critical, and may involve a fairly long trial-and-error phase.

d) **Security Concerns**:
- **Potential Vulnerabilities**: Fractal Encryption, which uses combinations to accomplish better transformation, might also remain relatively susceptible to cryptanalytic attacks. Since security threats are constant and new ones emerge frequently, evaluating the security measures and making improvements is required.

e) **Limited Research**:
- **Emerging Technology**: Unfortunately, Fractal Encryption is a relatively new technique and, thus, more research has not been done toward it than toward usual approaches. There is relatively limited published work on R-S as of 2018, meaning that some aspects of its current and long-term security and effectiveness in different settings may not be quite well understood.

## Conclusions

In this paper, modern innovations and optimisation methods are integrated to develop a comprehensive fractal encryption technique for securing image data. Fractal encryption and the Fourier transform are combined in an attempt to identify challenges of security and computation efficiency for image encryption. The comparative analysis of the proposed system expresses its superiority over classical cryptography algorithms. This is evident in encryption and decryption times, image fidelity, and computational efficiency. Encryption time is significantly reduced with a high level of security. These features make the proposed method suitable for real-time applications. In addition, the integrity and quality of the image are maintained, ensuring that the decrypted image remains uncorrupted.

## Future Works

Future studies must aim at streamlining the suggested fractal encryption architecture to minimize computation power and processing time even further by conducting parallel computations and hardware acceleration with the help of GPUs or specialized processors. Beyond that, more sophisticated and more adaptive fractal transformations could also be explored to maximize image fidelity, especially for large and complicated images, and permit dynamic parameter setting of the algorithm according to image content. Security-wise, a lot of testing of cryptographic robustness measures needs to be conducted to examine resistance to a variety of attack models, and hybrid encryption systems combining fractal techniques with chaos-based or contemporary symmetric cryptography can also help enhance protection. Lastly, generalization of the framework to other data modalities like video and multidimensional data, in addition to the real-world implementation and future performance assessment of the system, will give more insight into the scalability, stability, and usability applicability of fractal encryption systems.

## ACKNOWLEDGEMENT

The authors would like to show their heartfelt thanks to the University of Technology - Iraq, and Polytechnic College of Engineering-Baghdad, at the Middle Technical University (MTU) - Iraq, for supporting the present research by way of granting the required academic resources and facilities

## Conflict of Interest

The authors indicate that they have no conflict of interest as far as the publication of the current paper is concerned.